\documentclass[10pt]{iopart}
\usepackage{float}
\pdfoutput=1
\usepackage{parskip}
\usepackage{xcolor}
\usepackage{graphicx}
\usepackage{dcolumn}
\usepackage{bm}
\usepackage{amssymb}
\usepackage{blindtext}
\usepackage{hyperref}
\usepackage{parskip}
\usepackage{hyperref}
\usepackage{xcolor}
\usepackage{hyperref}
\usepackage{xcolor}
\newcommand{\orcidn}[1]{\href{https://orcid.org/#1}{\includegraphics[width=8pt]{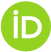}}}
\usepackage{multicol}
\usepackage{xcolor} 
\usepackage[export]{adjustbox}

\begin{document}

\title{Magnetic levitation by rotation  described by a new type of Levitron}

\author{A. Doff $^1$\orcidn{0000-0002-7808-0421} and R. M. Szmoski $^2$\orcidn{0000-0002-0968-7158} }

\address{$^{1,2}$ Universidade  Tecnol\'ogica  Federal do Paran\'a - UTFPR, R. Doutor Washington Subtil Chueire, 330 - Jardim Carvalho,  84017-220
Ponta Grossa,  DAFIS, Brazil}

\ead{ $^1$agomes@utfpr.edu.br,  $^2$rmszmoski@utfpr.edu.br}


\begin{abstract}
Recently, a novel magnetic levitation phenomenon involving two magnetically equivalent neodymium permanent magnets has been reported. In this work, we propose that this system functions as a scaled-up analog of the Levitron. The key distinction is that the ratio $m_/\mu_f$ becomes a function of the lateral displacement $\delta_R$, and magnetic trapping no longer depends on the rotational speed of the levitating body as in a conventional Levitron. Furthermore, we demonstrate that stable trapping occurs when a specific constraint on the $\delta_R$ parameter is satisfied, ensuring that the potential energy reaches a minimum at the equilibrium point.
\end{abstract}

\section{Introduction}

Magnetic levitation is a phenomenon in which an object is suspended using solely
magnetic forces to counterbalance gravitational force. This effect has important applications  in advanced technologies such as maglev trains and magnetic bearings. It can also be
demonstrated through simpler devices, such as the Levitron, commonly found as a toy.  In this, magnetic levitation is achieved through the repulsion between a rotating magnet and an external magnetic field. To maintain stability, a delicate balance must be achieved between  angular velocity ($\omega$) and the ratio of the spinner’s mass ($m$) to its magnetic momentum ($\mu$).  This balance allows the magnetic forces to counteract the downward force of gravity

As demonstrated in \cite{ref3} and \cite{ref4}, the effect of spin-stabilized magnetic levitation is a macroscopic analog of magnetic gradient traps used to confine particles possessing a quantum magnetic moment. Taking into account this analogy, the potential energy attributed to the levitron can be written as\cite{ref3}

\begin{eqnarray}
&& U(r,z) \approx \mu(B_0 + (\frac{mg}{\mu} +\frac{\partial B_z}{\partial z})z + \frac{1}{2}\frac{\partial^2 B_z}{\partial z^2}z^2 + \nonumber\\
&& \,\,\,\,\,\,\,\,\,\,\,\,\,\,\,\,\,\,\,\,\,\,\,\,\ \frac{1}{2}\frac{\partial^2 B_z}{\partial z^2}\left(\frac{1}{2B_0}\frac{(\frac{\partial B_z}{\partial z})^2}{\frac{\partial^2 B_z}{\partial z^2}} -1\right)r^2 +...)
\label{eq1}
\end{eqnarray}
\par Consequently, the condition for stable levitation can be described in terms of the ratio of the minimum to maximum spin frequency, which is given by \cite{ref3}

\begin{equation}
 \left(\frac{2r_{e} mg}{B \mu}\right) \approx \frac{\omega_{min}}{\omega_{max}} \approx O\left(\frac{1}{3}\right)
\label{eq2b}
\end{equation}

\noindent where $r_{e}$ is the effective radius for the moment of inertia
of the spinning top($I=mr^2_{e}$) and $B$ corresponds to the intensity of the applied magnetic field at the trapping position of the top.  Note that in Eq.(\ref{eq2b}) the effects of air resistance and other losses were not considered.

\par  Typically, in this type of device, the observed rotation frequencies range from 1,000 to 3,000 rpm, with  floating times of up to  4 minutes in air.

\par The magnets used in such devices are ring-shaped ferrite magnets(F), whose maximum intensity  is  of the order $B^F_{0} \sim O(0.4)T$. The main characteristic of neodymium magnets(N) is their high residual magnetic field intensity and coercivity\cite{ref6}, and  these magnets produce magnetic fields of the order $B^N_{0} \sim O(1.2)T$. 

\par Thus, we examine how the ratio given by Eq.(\ref{eq2b}) for a spinning top and magnetic base changes, considering it identical in dimensions to \cite{ref3} but composed of neodymium magnets.

\par In this case, it is possible to estimate $\left(\frac{\mu}{m}\right)_N = O(120)Am^2/Kg$ (see table IV \cite{ref2}), comparing with $\left(\frac{\mu}{m}\right)_F = 30.4 A m^2/Kg$\cite{ref3} and assuming the appropriate rescaling of $B_F$\cite{ref3} $\to B_N$\cite{ref6} we obtain for this hypothetical spinning top of neodymium

\begin{equation}
\left(\frac{2r_{e} mg}{B \mu}\right)_{N}  \approx \frac{\omega_{min}}{\omega_{max}} \approx  O\left(\frac{1}{25} \right). 
\label{eq2b2}
\end{equation}

\par Therefore, for the Levitron's rotation rate $\omega_{min} \approx O(500)$rpm\cite{ref7}, 
the rotation frequencies of this hypothetical top could range from  $500$ to $12,500$ rpm.
This rough estimate illustrates how the angular rotation of the magnet incorporated into the spinning top  would need to be modified to maintain the effect of spin-stabilized magnetic levitation in the presence of strong magnetic fields. 

Recently, in Refs.\cite{ref2}\cite{ref1},  the authors have proposed the existence of a novel type of magnetic levitation that uses two permanent magnets of similar size. One magnet, referred to as the rotor(R), is mounted on a motor with its north and south poles oriented perpendicularly to the axis of rotation, and it is set to rotate at angular velocities on the order of $10,000$ rpm

The second magnet, termed the floater(f), is placed in the vicinity of the rotating magnet: it is set in motion and it levitates toward the rotor until it floats in space a few centimeters below it. Note that the floater has a lower rotation frequency compared to the rotor, and  if perturbed, it experiences restoring forces that bring it back to its equilibrium position. 

The characterization of this phenomenon as a new type of magnetic levitation, as suggested in\cite{ref1},  arises  from observation that the spinning rate of the Levitron is $\sim O(500) rpm$\cite{ref7}, much lower than that seen  in this device.  

Note that the magnets used in the device described above are neodymium magnets.  As discussed previously, these magnets result in a significantly different angular speed range for the spinning top, potentially ranging from $500$ to $12,500$ rpm. Consequently, the rotation speed of the floater should be compared to the spinning top speed  suggested above.

\par Equation (\ref{eq2b}), derived for a conventional Levitron, demonstrates that the observed levitation effect is realized only when the ratio of mass to magnetic moment
$m_/\mu$ falls within the range of spinning top rotation frequencies required for the potential energy to reach a minimum at the trapping point.

\par   As we will demonstrate, in the system that we will refer to as the  "dynamical  Levitron" , which was considered in Refs.\cite{ref2}\cite{ref1}, the ratio of the mass to the magnetic moment $m_/\mu_f$ of the float is controlled   by the parameter $\delta_R$, and in this case the equation analogous to $(\ref{eq2b})$, corresponds to
 
 \begin{equation}
\left(\frac{2h mg}{B \mu_f}\right) \approx \frac{15}{4}\frac{\delta^c_{R}}{\delta^{max}_{R}},
\label{eqrdl}
\end{equation}

\noindent where $h$ is the height of a cylindrical neodymium magnet incorporated into the rotor, $\delta^c_R$ corresponds to critical value of $\delta_R$ necessary  to obtain the trapping of the float and finally $\delta^{max}_{R}$ corresponds to the maximum $\delta_R$  value. 

\par  Equation (\ref{eqrdl}) arise from the condition in which the floater presents the minimum energy required for trapping. Comparing Eqs.(\ref{eq2b}) and (\ref{eqrdl}), we can then suggest that the novel type of magnetic levitation mentioned in\cite{ref2}\cite{ref1} can, in fact, be interpreted as magnetic levitation resulting from a new type of Levitron.

\par Unlike Eq.(\ref{eq2b}), in the case of the dynamical Levitron the adjustment  of the ratio $m_/\mu_f$ is a function of the lateral displacement $\delta_R$ incorporated into the magnet coupled to the rotor. This means that the adjustment required for magnetic trapping is not determined by the floater (or the spinning top in the case of the conventional Levitron, considering a comparison  for illustrative purposes). Consequently, in this type of magnetic trapping it is possible to observe a situation in which the floater remains stationary.

\par  In Refs.\cite{ref3}\cite{ref4}, the authors established the fundamental physical principles underlying the magnetic levitation observed in conventional Levitron systems. In such systems, stable levitation requires a precise tuning of the radius $m_/\mu_f$, which in turn determines the allowable range of rotation frequencies for the spinning top, ensuring that the system reaches a potential minimum at the trapping point.
\par  In this work, we extend that framework to the case of a dynamical Levitron, demonstrating that the effective radius $m_/\mu_f$ becomes modulated by  $\delta_R$. This modulation provides a natural explanation for the magnetic levitation by rotation
described in Refs.\cite{ref2}\cite{ref1}. Furthermore, it allows us to identify the necessary conditions under which the system evolves toward a minimum energy configuration, thereby enabling stable magnetic trapping. Section II presents the  equations that describe the motion of the  floater, in section III based on considerations about the system's potential energy, we establish the necessary conditions to explain the magnetic trapping effect. Sections IV and V  details our results, which are organized into five subsections . Finally in section VI we present our conclusions.

\section{ Neodymium floater dynamics}

In order to characterize the dynamics of this system, the floater (f) 
can be reasonably approximated as a magnetic dipole  whose center coincides with the center of mass \cite{ref1}\cite{ref3}. The positions of the center of mass and the dipole are located at the same coordinates $\vec{r}$, subject to the action of the external field $\vec{B}_R$ produced by the rotating magnet(R). Therefore, the  equations that describe the motion of the  floater are\cite{ref1}
    \begin{eqnarray}
        && I_f\frac{d\omega_f}{dt} = \vec{\mu}_f \times \vec{B}_R - \xi^R_{f}\omega_f\nonumber \\
        &&  m_f\frac{d^2\vec{r}}{dt^2} = \vec{\nabla}(\vec{\mu}_f.\vec{B}_R) - mg\hat{z} - \xi^T_{f}v_f
        \label{eq5}
    \end{eqnarray}
\noindent where in the above expression $I_f$ corresponds to the moment of inertia of the floater, $v_f$ , $m_f$  and $\mu_f$ correspond respectively to the velocity, mass and magnetic momentum, $g = 9.81 m/s^2$ is the gravitational acceleration and $\vec{B}_R$ is the rotor field evaluated at the center of the floater. 

The respective drag coefficients associated with the rotational and translational displacements of floater are indicated by  $\xi^R_{f}$, $\xi^T_{f}$ and  in these equations we can identify
the dipole force from the rotor on the floater as 
\begin{equation}
\vec{F}_{dip} = \vec{\nabla}(\vec{\mu}_f.\vec{B}_R),
\label{eq6}
\end{equation}

\noindent where $U_f = -\vec{\mu}_f.\vec{B}_R$ . In the absence of air resistance, or in the limit where $\xi^R_{f} = \xi^T_{f} = 0$, Eq.(\ref{eq5}) corresponds to equations (1) and (2) of Ref.\cite{ref3} . In these equations, we will assume that the  dynamics of the float is similar to that of a spinning top\cite{ref3}  composed of  neodymium magnets.

The observed interaction forces  between the rotor(R) and the floater(f), which are modeled by the dipolar approximation given by Eq.(\ref{eq6}), can be expressed using a Taylor expansion as\cite{ref1}

\begin{equation}
\vec{F}_{dip} \approx \frac{3\mu_0\mu_f\mu_R}{4\pi r_{f}^4}\left(\frac{4\delta_R}{r_f} - R_f\right) 
\label{eq7}
\end{equation}

\noindent where in the  above equation  $\mu_0 = 4\pi \times 10^{-7} T.m/A$, $r_f$ represents  the distance of the magnetic rotor in relation to the float and $\mu_{R,f}$ are the magnetic momentum of the rotor and floater. 

As reported in\cite{ref1} the parameter $\delta_R$ incorporates a small lateral shift away from the rotation axis, while  $R_f = \frac{\omega_f}{\omega_R} = \sin\theta_f$  and $\theta_f$ is the polar angle that is associated with the angular velocities of the floater and rotor.

 \begin{figure}[t]
\vspace*{-1cm}
    \centering
\includegraphics[scale=0.4]{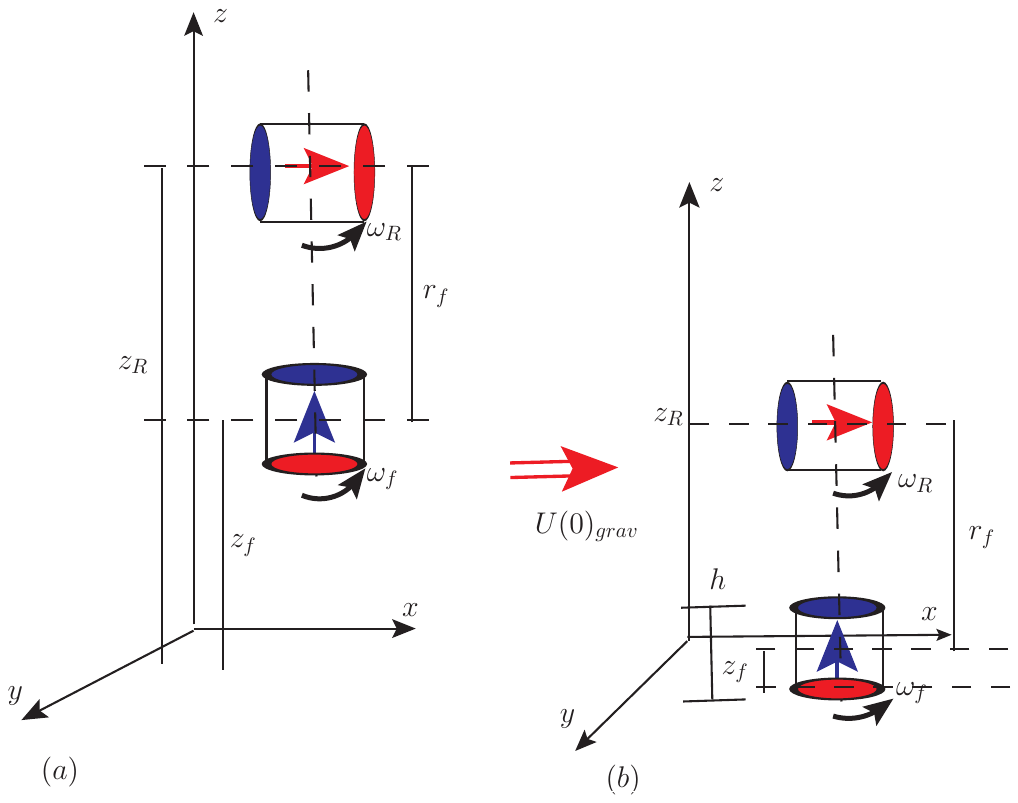}
\vspace*{-5.2cm}
\caption{ FIG.(1a) depicts the coordinates of the rotor (R) and floater (f) relative to $r_f = z_R - z_f$. The red arrow illustrates the  orientation of the magnetic moment attributed to the rotor $\mu_R$, while the blue arrow represents the corresponding magnetic moment of the floater $\mu_f$.  FIG. (1b),  represent the scheme for characterizing the float’s energy as a function of the coordinates $(z_f, z_R)$, which define the ground state $U(0)_{grav}$ of the float.
     }
    \label{fig1}
\end{figure}

\section{ The Potencial Energy}

Assuming Eqs.(\ref{eq6}) and (\ref{eq7}), we can represent the potential energy between the rotor and the float, in order to compare the behavior of this system with that of the Levitron discussed in\cite{ref3}, in the following form
 
 \begin{equation}
U_{f}(z_f)   \approx -\frac{\mu_0\mu_f\mu_R}{4\pi}\left(\frac{3\delta_R}{(z_R - z_f)^4} - \frac{R_f}{(z_R - z_f)^3}\right) +  mgz_f. 
\label{eq8.1}
\end{equation}

\noindent where $r_f = z_R - z_f$ and  $z_{f,R}(mm)$ corresponds to the coordinates of the rotor and float in the vertical direction represented in Fig.1.

\par As discussed the effect of spin-stabilized magnetic levitation observed in the Levitron provides a macroscopic analogue of magnetic gradient traps used to confine particles with a quantum magnetic moment.

\par In this case,  under suitable conditions, the component of the magnetic moment along the local magnetic field direction is an adiabatic invariant and the gyroscopic action must
not only prevent the floater from flipping, but also continuously align
the floater's precession axis along the local magnetic field direction.

\par It follows that, the observed levitation effect is only achieved when the ratio of the mass to the magnetic moment $m_/\mu$ yields the range for the spinning top rotation frequencies given by (\ref{eq2b}),  where this ratio is adjusted by adding small weights to  the spinning top. 

\par  For the rotor(R) and floater(f) system, the necessary correction is mediated by the parameter $\delta_R$ as illustrated in  Eq.(\ref{eqrdl}), which controls the adjustment of the float  precession axis along the magnetic field direction.

\par Therefore, the $\delta_R$ parameter  will be responsible for controlling the levitation effect of the floater.  The necessary conditions to observe this effect can be summarized as follows:

\par 
\vspace*{0.3cm}
\par\noindent {\it (i) The potential energy must have a local minimum at the trapping point, where
 $U_{z}(z_f) > 0$}  \cite{ref3}\cite{ref4}
\par\noindent {\it (ii) Additionally, vertical stability requires that the second derivative satisfies $\frac{\partial^2 U_{z}(z_f) }{\partial z_{f}^2} \equiv \partial^2_{z}U_z > 0 $\cite{ref3}\cite{ref4}}.

\par Furthermore, it is necessary that the system under the trapping condition exhibits dynamical equilibrium ,  i. e.  $\vec{\nabla U}_z = 0$, along with horizontal stability, i.e. $\partial^2_{x,y}U > 0$ \cite{ref3}\cite{ref4}. For simplification purposes, we assume that, under observation of trapping,  the system presents  horizontal stability.

\par   Once the necessary conditions for trapping, we can now consider the following situations. 
In the case that we can refer to as decoupled, where $\delta_R \approx 0$, which can be seen as a condition below a critical value $\delta_R < \delta^c_R $, the system will not present the trapping  conditions given by  (i) and (ii). 

\par On the other hand, for the situation that can be designated as coupled, where we typically observe  $\delta_R  \gtrsim \delta^c_R$, system begins to satisfy the conditions described in (i) and (ii), where the factor $\delta^c_R$  corresponds to a minimum (or critical) value required for  stable levitation. In the following sections, we will consider separately the situations in which $\delta_R <  \delta^c_R$(decoupled) and  $\delta_R  \gtrsim \delta^c_R$(coupled).

\section{The decoupled case}

\begin{figure}[t]
    \centering
\hspace*{-0.5cm}\includegraphics[scale=0.5]{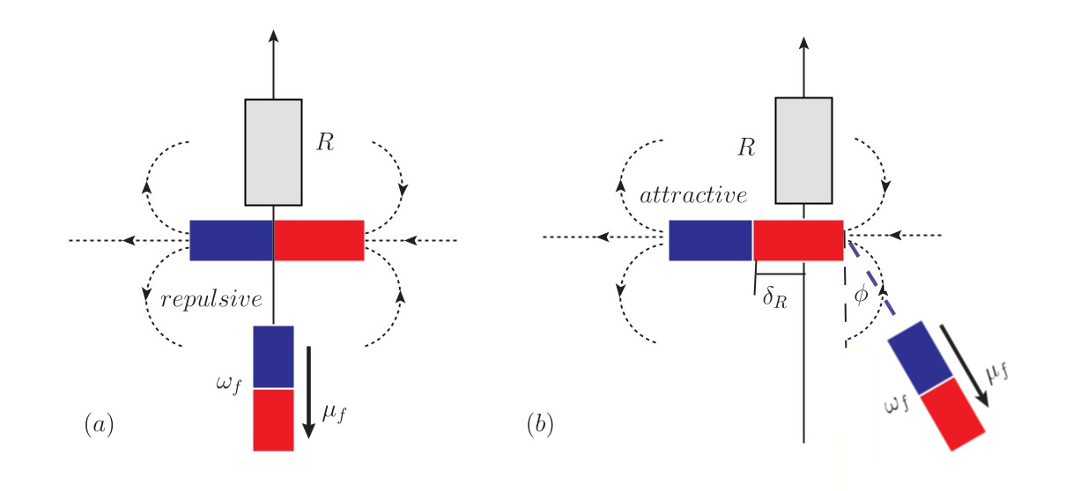}
 \vspace*{-0.1cm}
    \caption{Representation of the configuration of the magnets coupled to the rotor (R) for situations in which the lateral displacement $\delta_R$ corresponds to (a) $\delta_R = 0$ and (b) $\delta_R > \delta^c_R$. }
    \label{fig2}
\end{figure}

\par  FIG.(2a) shows the arrangement between the rotor (R) and the floater(f)  in a scenario with a high degree of symmetry  with respect to the z-axis, such that we can consider $\delta_R = 0$ . At this point, we highlight the differences between  a conventional Levitron\cite{ref3}\cite{ref4},  and the so-called  the "dynamical  Levitron". 

\par In the dynamical case, there is a transfer of angular momentum from the base, which now rotates, to the float  mediated by the configuration of magnetic field lines determined by the positioning of the magnet coupled to the rotor. Due to
 the effects of dynamical drag forces, it is expected that  $\omega_f < \omega_R$.

\par In a conventional Levitron, the gyroscopic effect is characterized
by the ratio  $\omega_{min}/\omega_{max}$. As we previously noted, this ratio is not only responsible for preventing  the floater  from flipping ,  in order to maintain the stability of the trapping, but it also ensures the continuous alignment of the floater's precession axis with the direction of the local magnetic field.

\par In contrast, for the dynamical Levitron, the parameter $\delta_R$ is responsible for the alignment  incorporated into the arrangement considered for fixing the magnet to the rotor as illustrated in FIG.(2b). Importantly,  even in the limit where $R_f \to 0(or\,\, \omega_f\to 0) $, as we will see in the next section, the necessary trapping conditions (i) and (ii) still apply.

\par Considering these fundamental distinctions between the two systems, and incorporating the condition required for dynamic equilibrium, we can express Equation (\ref{eq8.1}) as follows

\begin{eqnarray}
&& U_{f}(z_f)   \approx -\frac{\mu_0\mu_f\mu_R}{4\pi}\left(\frac{3\delta_R}{(z_R - z_f)^4} - \frac{R_f}{(z_R - z_f)^3}\right) + \nonumber\\
&&\,\,\,\,\,\,\,\,\,\,\,\,\,\,\,\,\,\,\,\,\,\,\,\,\,\,\,\,\,\,\frac{\mu_0\mu_f\mu_R}{4\pi}\left(\frac{12\delta_Rz_f}{(z_R - z_f)^5} - \frac{3R_fz_f}{(z_R - z_f)^4}\right) . 
\label{eq81a}
\end{eqnarray}

\par Then, assuming the limit in which $\delta_R \to  0$, we obtain 

\begin{equation}
U_{f}(z_f)   \approx \frac{\mu_0\mu_f\mu_R R_f}{4\pi}\left(\frac{1}{(z_R - z_f)^3}  - \frac{3z_f}{(z_R - z_f)^4}\right) , 
\label{eq11a}
\end{equation}
 \noindent and the above expression shows that for  $\delta_R \to 0$,  that condition (i) can be satisfied  only if  $3z_f/(z_R - z_f) < 1$.

\par In addition to the possibility raised above, it is crucial to emphasize that condition (ii) must be satisfied. Consequently, in this limit, we obtain:

\begin{eqnarray}
&& \frac{\partial^2 U_{f}(z_f) }{\partial z_{f}^2} = -\frac{\mu_0\mu_f\mu_R}{4\pi}\left( \frac{12R_f}{(z_R - z_f)^5} +\frac{60R_fz_f}{(z_R - z_f)^6} \right).   \nonumber \\
\label{eq11b}
\end{eqnarray}

\noindent Therefore, even assuming that $3z_f/(z_R - z_f) < 1$, 
which is consistent with condition (i) ,  the above expression shows that
 the system is not able to satisfy both stability conditions simultaneously.

\par The discussion  in the previous paragraphs highlights the necessity for
a minimum, or critical value of $\delta_R=\delta^c_R$, required to achieve
the trapping of the float. The condition that characterize this critical value of the  lateral displacement,  assuming the limit where $R_f\to 0$, that corresponds to $\delta^c_R = \delta_R(0)$, is given in terms of the float's energy, which corresponds exclusively to its gravitational potential energy $U(0)_{grav}$.

\par  In order to estimate the required value of $\delta_R$ to satisfy the stability conditions, FIG.(3) presents a contour plot  of $U_{f}(z_f)$  and $\partial^2_{z_f}U_f(z_f)$, assuming   $\delta_R = 0.1mm$ for illustrative purposes. The y-axis of this figure represents $R_f$, while the x-axis represents  $z_f(mm)$. To establish an adequate scale for this plot, the term $\partial^2_{z_f}U_f$ is multiplied by $10^{-5}$.

\par  In  FIG(3a), it is evident that condition (i) is maintained, as indicated by the purple point. This region, depicted in blue, represents an area of stability, while the instability region is delimited in red. Therefore, as shown in FIG.(3b), for the same region in the $(z_f(mm),R_f)$ plane, condition (ii) is not simultaneously satisfied, and it can be verified  that $\delta_R < \delta^c_R$.

\par  To provide clarification the behavior observed in this figure,  it is advantageous to review FIG. (1).  FIG. (1b),  presents a schematic illustrating how we characterize the energy of the float as a function of the coordinates $(z_f, z_R)$, which define the ground state for the gravitational potential energy of the floater, denoted by $U(0)_{grav}$.

\par In the plane  shown in FIG.(1b), when $z=0$, the base of the cylindrical floater in this figure defines the coordinate $z_f = h/2$, whereas for a spherical floater $z_f = D/2$,  where $h$ is the height of the cylinder and $D$ is the diameter of a sphere. 

\par Therefore,   the minimum of gravitational potential energy is given by $U(0)_{grav} = mgz_f$.  Once we identify $r_f =( z_R - z_f) \,mm $,  the minimum choice for $z_R(mm)$, necessary  to contemplate $r_f \approx O(10)mm$ typically observed in experiments performed with spherical and cylindrical floats, would be $z_R \approx O(14)mm$.

\begin{figure}[t]
    \centering
\hspace*{-0.2cm}\includegraphics[scale=0.4]{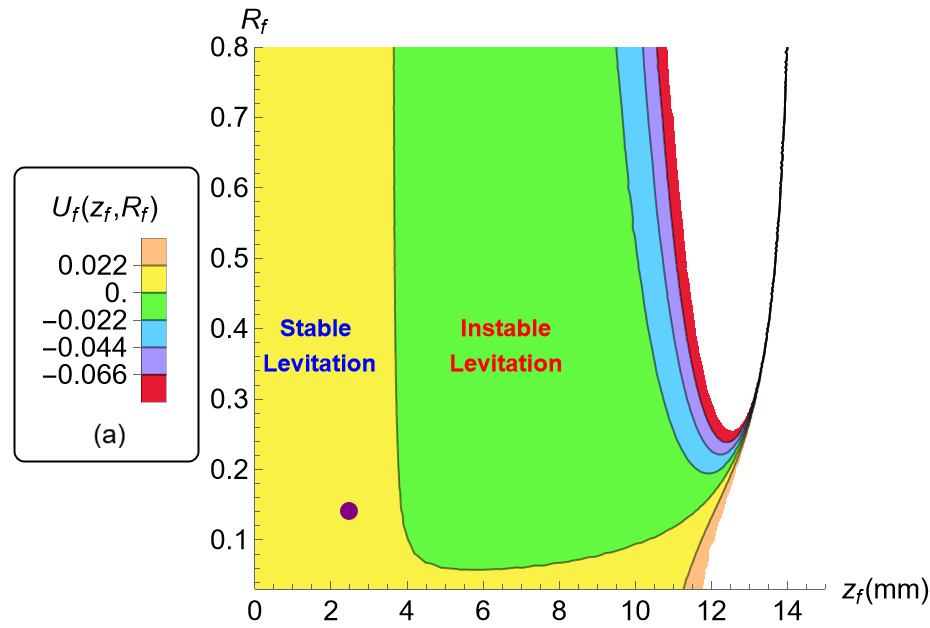}\hspace*{0.05cm}\includegraphics[scale=0.4]{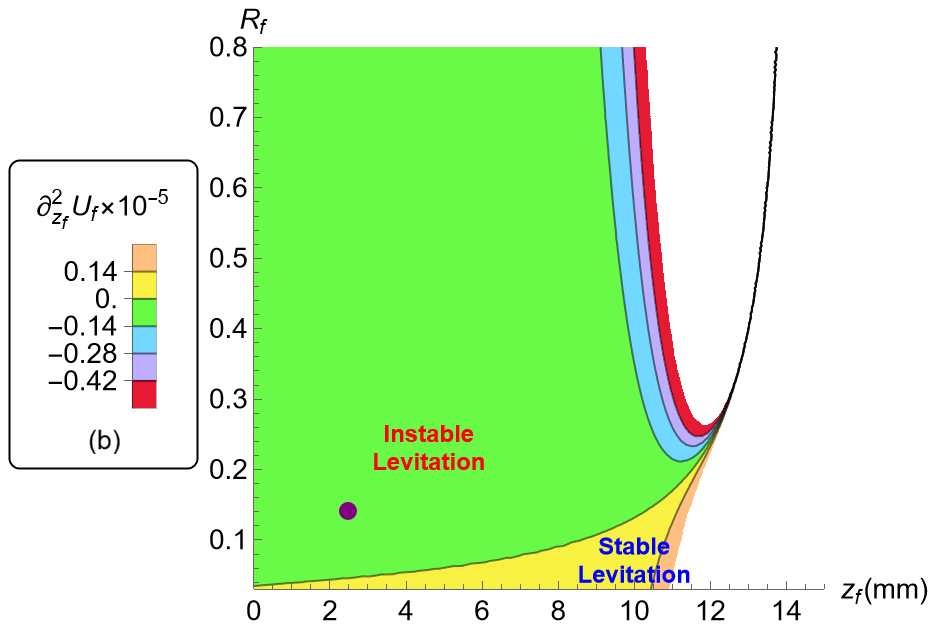}
    \caption{Contour plot obtained from Eq.(\ref{eq81a})  for $\delta_R = 0.1mm$. The contextualization of the curves behavior are described in the text.}
    \label{fig3}
\end{figure}

\par In FIG.(3) we consider the choice $r_f = (14.2 - z_f)\,mm$, which is adequate to represent the results obtained for floaters formed by cylinders of height $h=5$, or spheres of diameter $D=(4)7$  as listed in the tables presented in Appendix B.

\par Choices where $z_R > 14 mm$ lead to corrections of the form  $U(0)_{grav} + \delta U_{grav}$, and tend to decrease in plane  $(z_f(mm),R_f)$  towards regions where $U_z(z_f) < 0$ is observed. However, in addition to condition (i) the characterization of the critical behavior for the trap’s stability is determined  by the criterion $\frac{\partial^2 U_{f}(z_f) }{\partial z_{f}^2} > 0$,
which is independent of any constant added.

\par Finally, we can consider  the lateral displacement induced by $\delta_R$. 
As illustrated in FIG(2b) , this displacement introduces an attractive component to the dipolar interaction. Consequently, the magnetic interaction serves both to counterbalance the gravitational force on the floater and to align a component of its magnetic moment along the direction of the local magnetic field. As we will demonstrate in the next section, this configuration simultaneously satisfies the trapping conditions given by (i) and (ii).

\section{The coupled case}

\begin{figure*}[t]
    \centering
\includegraphics[scale=0.3]{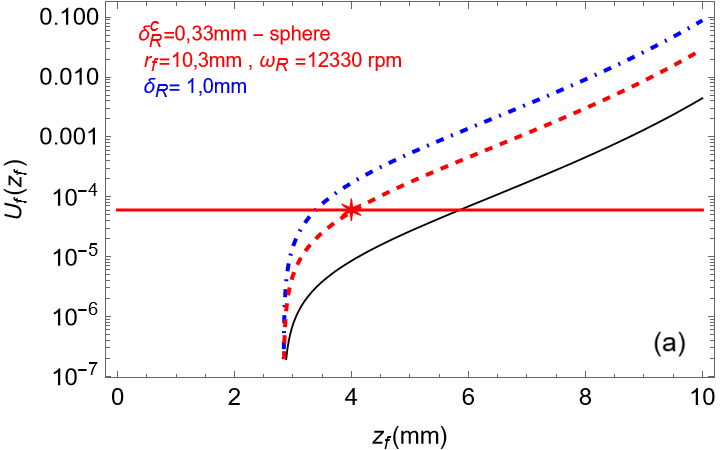}\hspace*{0.5cm}\includegraphics[scale=0.3]{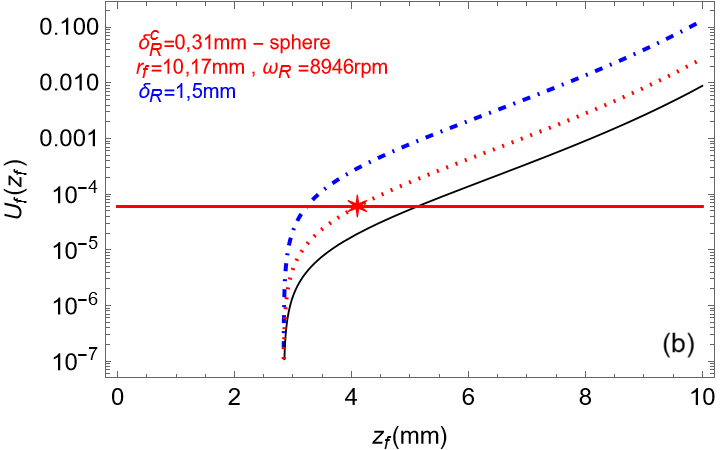}\hspace*{0.5cm}\includegraphics[scale=0.3]{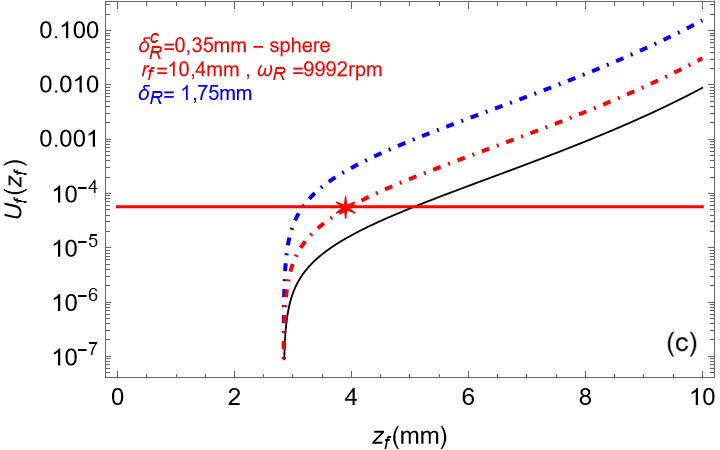}\\
\includegraphics[scale=0.3]{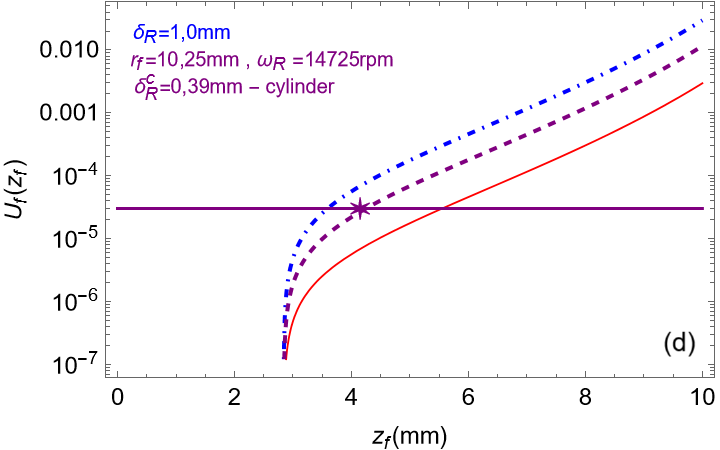}\hspace*{0.5cm}\includegraphics[scale=0.3]{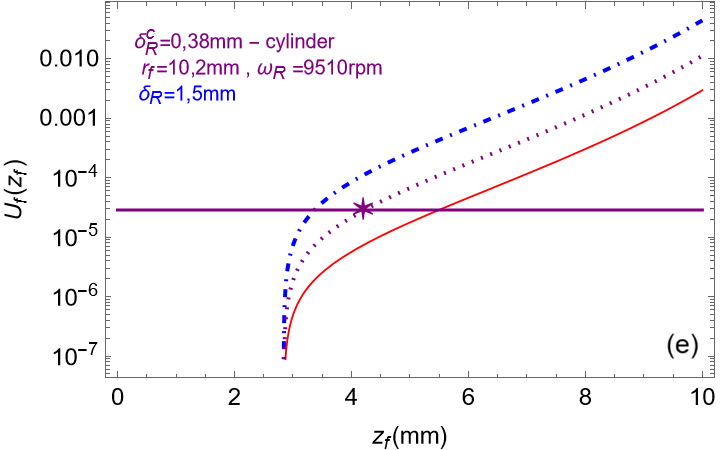}\hspace*{0.5cm}\includegraphics[scale=0.3]{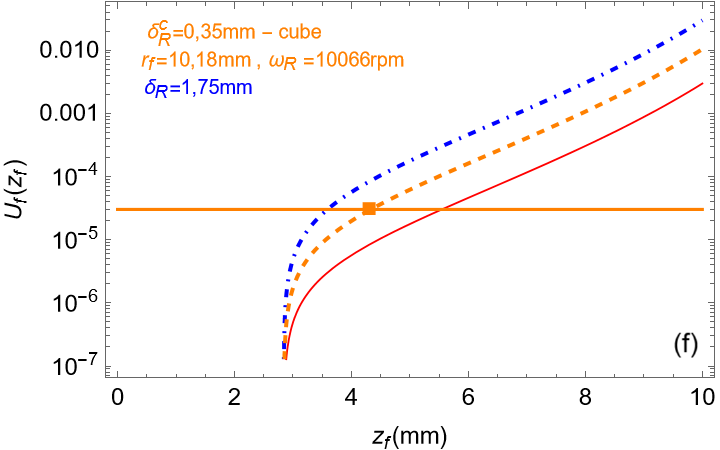}
    \caption{Plot obtained for Eq.(\ref{eq81a})  assuming $R_f = 0 $  in the cases where
     $\delta_R =1mm$, $1.5mm$ and $1.75mm$  for  spherical, cylindrical and cube floats represented by  blue dot-dash curves.  The dimensions and mass of the cylindrical and cube floats correspond to width and height of $5mm$,  $m_f = 0.73gr$, while for a spherical float the respective measurements are diameter $7mm$,  mass $m_f = 1.53gr$. The magnetic momentum of cylindrical(cube)  floats and rotor are $\mu_f = \mu_R = 0.063 A.m^2$, while that of the spherical is $\mu_f = 0.18  A.m^2$, the contextualization of the curves behavior are described in the text.}
    \label{fig4}
\end{figure*}

\subsection{ The minimum energy condition for the trapping: Floater at rest}

\par  FIG.(2b)  illustrates the configuration in which an asymmetry of the rotor axis with respect to the z-axis is observed. In this case, for  $\delta_R  > 0$, Eq.(\ref{eq11b}) can be written as

\begin{eqnarray}
&& \frac{\partial^2 U_{f}(z_f) }{\partial z_{f}^2} = \frac{\mu_0\mu_f\mu_R}{4\pi}\left(\frac{60\delta_R}{(z_R - z_f)^6} -\frac{12R_f}{(z_R - z_f)^5}\, + \right.\nonumber\\
&& \left.\,\,\,\,\,\,\,\,\,\,\,\,\,\,\,\,\,\,\,\,\,\,\,\,\,\,\,\,\,\,\,\,\,\,\frac{360\delta_R z_f}{(z_R - z_f)^7} -\frac{60R_f z_f}{(z_R - z_f)^6} \right),  \nonumber \\
\label{eq11d}
\end{eqnarray}

\noindent so that the addition of the terms proportional to $\delta_R$, which are positive, now allows us to obtain $\partial^2_{z}U_z > 0$. Therefore, 
under the configuration shown in FIG.(1b) , Eq.(\ref{eq11d}) yields the following condition on $\delta_R$, when $U_z(grav) = U(0)_{grav}$
for a given floater geometry

\begin{eqnarray}
&&\left(60\delta_R r_f -12R_fr_f^2 + 180\delta_R L - 30R_f Lr_f \right) > 0,  \nonumber \\
\label{eq11c}
\end{eqnarray}

\noindent  where $L = h$  for  cylindrical(or cube) floaters and $L = D$ for a spherical floater. 
 From the above expression, we can thus characterize the minimum or critical value $\delta^c_R$ required to satisfy the trapping conditions given by (i) and (ii).

\par  It can be verified, by inspection of Eq.(\ref{eq11c}), that the stability condition (ii) is still maintained in the limit where we can assume $R_f \to 0$, which then suggests, within the context of a dynamical  Levitron, the possibility of observing the trapping of the floater at rest. 

\subsection{\texorpdfstring{The experimental determination of $\delta^c_R$ for an floater at rest}{} }

\par The experimental setup employed in this work is thoroughly detailed in Appendix A. As described therein, our study involved the recording of the phenomenon and precise measurements of the rotor and floater rpm, utilizing a  digital tachometer and  high-resolution camera operating at 960 frames per second (fps). A rigorous analysis of these videos was subsequently performed using the Tracker software. All video recordings of the experiments conducted in this study, which document the experimental observations of trapping, are archived and made publicly available in the repository referenced in \cite{videos1}. Among the diverse conditions under which trapping was observed and recorded, we particularly highlight the possibility of trapping occurring even when the floater remains at rest. The quantitative results extracted from this and all other related analyses are systematically compiled in tables presented in Appendix B.

 \par Based on direct measurements of the distance ($r_f$) between the rotor and the floater, in FIG.(4)  we analyze the behavior of $ U_{f}(z_f)$ for $R_f = 0$ for spherical, cylindrical, and cubic floaters as a function of the lateral displacement $\delta_R$. For the spherical floater, the cases $\delta_R \approx 1mm$, $1.5mm$ and $1.75mm$ are shown in FIGs. (4a)-(4c), respectively. For the cylindrical floater, the cases $\delta_R \approx 1mm$, $1.5mm$  are illustrated in FIGs. 4(d) and 4(e), while FIG. 4(f) presents the results for the cubic floater, which has the same dimensions as the cylinder, with $\delta_R = 1.75mm$.

\par The set of plots presented in FIG(4)  summarizes three different configurations of magnets coupled to the rotor. In presenting these results, we use a consistent color scheme for the curves corresponding to each case: red for the spherical floater, purple for the cylindrical floater, and orange for the cubic floater.

\par In these plots, the red, purple, and orange points 
$({\bf\color{red}\ast}, {\bf\color{purple}\ast}, {\bf\color{orange}\blacksquare})$  correspond to the condition where the floaters presents $\omega_f=0$\footnote{Note that the floater initially oscillates harmonically in the (x,y) plane, $\theta_{(x,y)_{+}} \leftrightarrow \theta_{(x,y)_{-}}$ for only a short time interval, which defines $\bar{\omega}_f=\Delta \theta_{(x,y)}/\Delta t = 0$ after a period of time. It is possible to identify $\theta_{(x,y)_{+}}=\theta_{(x,y)_{-}} \to 0$ in the asymptotic limit.}. In this scenario,
 the floaters are characterized solely by their gravitational potential energy $U(0)_{grav}=mgz_f$, represented by the continuous red, purple or orange  lines. The position $z_f$  of the floater results from the direct measurement of $r_f$,  which is indicated in the upper left corner, observed for the trapping described in the videos\cite{videos1}.
 
 \par The blue dot-dashed curves,  derived  from Eq.(\ref{eq81a}) for the cases  $\delta_R = (1, 1.5, 1.75)mm$ and $R_f = 0$, illustrate scenarios where the lateral displacement exceeds the critical threshold $\delta_R > \delta^c_R$. In contrast, the black (or red) curves correspond to cases in which the lateral displacement remains below the critical threshold ( $\delta_R < \delta^c_R$).

\par  As shown in FIG.(4b), the red dotted line corresponding to $\delta_R =0.31mm$,
 intersects the point (${\bf\color{red}\ast}$) in $z_f\approx 3.9mm$ indicating
 that the minimum condition for trapping of floater occur when $\delta^c_R \approx O(0.3)mm$. Similar results are observed in the other figures; in FIG.(4a) we find $\delta_R =0.33 mm$,  while in FIG.(4c)  $\delta_R =0.35mm$ respectively, to $z_f\approx 4 mm$. For the cylindrical floater, the results are also comparable. In FIGs.(4d),(4e),  we obtain  $\delta^c_R \approx 0.39mm$ and  $\delta^c_R \approx 0.38mm$, while for the cubic float  in  FIG.(4f) we find  $\delta^c_R \approx 0.35mm$.

\par The  experimental results presented in FIG.(4), obtained for different floaters and lateral displacements $\delta_R$, yield closely matching estimates for the critical value $\delta^c_R$.  These findings suggest, that the minimum value $\delta^c_R$ required for trapping is of the order of $\sim O(0.3-0.4)mm$.  Furthermore, they demonstrate that the magnetic alignment effect required for trapping is  effectively achieved by a suitable choice of $\delta_R$, regardless of the geometry of the floater.

\par  The results found for $\delta_R$ are in agreement with the Eq.(\ref{eq11c}), which establishes that in the limit where $R_F \to 0$,  that the trapping condition  (ii) can only be satisfied for $\delta_R > 0$. However, it is important to emphasize that in this limit Eq.(\ref{eq11c}) is not able to predict a specific value for $\delta^c_R$, with $\delta^c_R$ being defined by the minimum energy of the float required to maintain trapping, where the energy of the float $E_f$ corresponds to $U(0)_{grav}$. 

\par In this work we conducted a series of measurements, varying  $\delta_R = (0.1, 0.3, 1, 1.5, 1.75,2.5)mm$, and verified that float trapping actually begins at $\delta_R \approx O(0.3)mm$. Additionally, the stability of the trapped floater persists for a considerable time when  $\delta_R  \gtrsim  1 mm$.  The critical value $\delta^c_R \approx O(0.3-0.4)mm$ obtained  from  FIG.(4) is consistent with our observations for the magnetic trapping  in several measurements.

\par  Depending on the specific experimental setup, $\delta^c_R$ is expected to depend on the shape (or geometry) of the magnet affixed to the rotor. This dependency arises because each configuration of the magnet will result in a distinct configuration of magnetic field lines, necessitating continuous adjustments to align the floater precession axis.

\par   The results reported in Figs.(4a-4f) are from floaters with diverse geometries. Consequently, for a more precise determination for $\delta^c_R$ it would be necessary to characterize the  non-uniformity of magnetic moments inherent for example to cylindrical and cubic floaters that would go beyond the scope of this work, which aims to establish the basic conditions necessary to explain the observed magnetic trapping.

\par As noted in Ref.\cite{ref1}, the trapping effect is observed only when $\delta_R > 1mm$, where the simulations reported by those authors utilized a $\delta_R$ value of 1.2mm. Within our experimental setup, the magnet coupled to the rotor consists of a neodymium cylinder with both height and diameter $d=h=5mm$, whereas the magnet used in the experiments detailed in \cite{ref1} corresponds to a neodymium sphere with a diameter of $D=19mm$.

\par Applying a straightforward dimensional analysis, we estimate that the
 critical value for the experimental apparatus employed in \cite{ref1} would be of the order $\delta^c_R (19mm) \approx \frac{D}{h}\delta^c_R(5mm) \approx O(1)mm$  which agrees with our  observations and is also in accordance with recent results reported in \cite{ref9}  highlights the sensitivity of $\delta^c_R$ to the geometry of the magnet.

\begin{table}[ht]
\caption{Experimental measurements arising from Tables (B1-B6)  for $r_f(mm)$  and $R_f$, with  $\delta^c_{R}$ given by Eq.(\ref{eqs}), for  cylindrical and spherical  floats  with diameter and height of $D_1=7mm$, $D_2= 4mm$ and $h=d =5mm$ . }
\centering
\begin{tabular}{ccccc}
\hspace*{1cm}\textrm{Float}&
\textrm{$\delta_R(mm)$}&
\textrm{$r_f(mm)$}&
\textrm{$R_f$}&
\textrm{$\delta^c_R(mm)$}\\
\hline
  &   &  9.4$\pm$ 0.094 & 0.12$\pm$  0.0012 & 0.23$\pm$ 0.023 \\
\hspace*{1cm}cylinder & 1.5mm & 9.6$\pm$ 0.096  & 0.115$\pm$ 0.0011 & 0.22$\pm$ 0.022\\ 
 & & 10$\pm$ 0.1 & 0.11$\pm$ 0.0011 & 0.22$\pm$ 0.0022  \\ 
   & 1.75mm & 9.7$\pm$ 0.097  & 0.162$\pm$ 0.0016 & 0.314$\pm$ 0.0031\\
\hline 
 &  & 9.52$\pm$ 0.095  & 0.134$\pm$ 0.0013 & 0.26$\pm$ 0.0026 \\
\hspace*{1cm}sphere $(D=7mm)$& 1.5mm  & 9.74$\pm$ 0.097 & 0.125$\pm$ 0.0012 & 0.243$\pm$ 0.0024 \\ 
  &  & 9.9$\pm$ 0.099 & 0.12$\pm$ 0.0012 & 0.24$\pm$ 0.0024\\
\hline 
& &  9.76$\pm$ 0.098  & 0.134$\pm$ 0.0013 & 0.26$\pm$ 0.0026 \\
\hspace*{1cm}sphere $(D=4mm)$& 1,5mm  & 10.22$\pm$ 0.1 & 0.132$\pm$ 0.0013 & 0.27$\pm$ 0.0027 \\ 
 &  & 11.26$\pm$ 0.11  & 0.11$\pm$ 0.0011& 0.248$\pm$ 0.0025 \\
 \hline
 &  & 9.22$\pm$ 0.092  & 0.145$\pm$ 0.0014 & 0.267$\pm$ 0.0027 \\
\hspace*{1cm}sphere $(D=7mm)$& 1.75mm  & 9.60$\pm$ 0.096 & 0.129$\pm$ 0.09013 & 0.248$\pm$ 0.0025 \\ 
  &  & 10.03$\pm$ 0.1 & 0.113$\pm$ 0.0011 & 0.23$\pm$ 0.0023\\
 \hline
\end{tabular}

\end{table}

\par In the following subsection, based on the series of measurements summarized in the tables of Appendix B, we will determine $\delta^c_R$  from Eq.(\ref{eq11c}) assuming $R_F > 0$, which allows in this case to extract $\delta^c_R$ directly from Eq.(\ref{eq11c}) in order to compare with the results presented in this section.

\subsection{  The minimum energy condition  for a dynamical  trapping: The minimum rotational energy of the floater }

\par In the previous section, we established the conditions for trapping a floater at rest, defined by $R_f \to 0$. We found that the minimum value of energy $U_f(z_f)$, required to maintain this trapping, demands a minimum value assigned to the lateral displacement $\delta_R$, defined by $\delta^c_R \approx O(0.3 - 0.4)mm$. Now, assuming that floater has the minimum rotational energy
we will, based on experimental measurements for $R_f$, verify the agreement of the predictions of Eq.(\ref{eq11c}) with the results previously obtained.

\par Assuming that the floater is  trapped, but now with angular velocity $\omega_f > 0$,  Eq.(\ref{eq11c})  yields
\begin{eqnarray}
&& \delta_R > \frac{(r_fR_f)}{10}\frac{(2r_f +5L)}{(r_f +3L)}. 
\label{eq11e}
\end{eqnarray}

\noindent For $R_f > 0 $, in Eq.(\ref{eq11e}) the value assigned to $R_f \to (R_f)_{min}$ corresponds to the situation where the floater has the minimum rotational  speed while being trapped, which consequently approaches the minimum value for $\delta_R$. At this point, the distance between the rotor and the floater $r_f$ is maximized approaching   distance defined for $\delta^c_R$ when $R_f \approx 0$. Assuming the limit  $r_f >> L$, it follows that
 
\begin{equation}
   \delta^c_R \gtrsim  \frac{r_fR_{f}(min)}{5}.
   \label{eqs}
\end{equation}

\subsection {\texorpdfstring{The experimental determination of $\delta^c_R$ for an dynamical  trapping }{}}

\par Based on the experimental measurements of $r_f$ and $R_f$ extracted from Tables $B1-B6$, we directly determine $\delta^c_R$ using  Eq.(\ref{eqs}) . In Table 1, the calculated $\delta^c_R$ values range from $\delta^c_R = (0.23-0.31)$mm , which are in good agreement with previous findings. It is important to note that Eq.(\ref{eqs})  is derived under the point-dipole approximation for the floater. Finite-size corrections, parameterized by the multiplicative factor $(2 +5L/r_f)/(1 +3L/r_f)$, reduce the prediction of this equation by a factor $ \lesssim O(0.1)mm$. 

\par Having established a reasonable agreement between the $\delta^c_R$   values obtained directly from Eq.(\ref{eqs}) and previous results, we can now proceed to characterize the necessary conditions for the minimum rotational energy of the floater, denoted as $R_{f}(min)$. Utilizing the results from the previous section, specifically  $\delta^c_R \sim (0.3-0.4)$mm
, and the observation that the maximum trapping distance for the floaters corresponds to $r_f \approx (10-11)$mm, we then obtain 

\begin{figure}[t]
    \centering
     \includegraphics[scale=0.5]{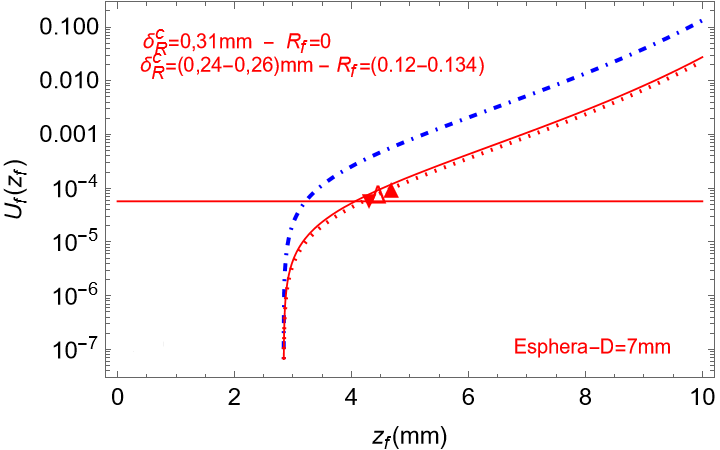}  
    \caption{Plot obtained for Eq.(\ref{eq81a})  assuming $(R_f)_{min}$ for  spherical float considering the data in Table 1 for $\delta_R=1.5mm$. The contextualization of the curves behavior are described in the text.}
    \label{fig5}
\end{figure}

\begin{equation}
   R_{f}(min)  \lesssim 0.17.
   \label{eqs2}
\end{equation}

\par Based on above expression,  the rotational speed of the floater relative to the rotor,  where $r_f \approx O(10)mm$, should satisfy $\omega_f \lesssim 0.17 \omega_R$. The prediction $\omega_f \lesssim 0.17 \omega_R$ is corroborated by the results reported in Tables (B1-B4) for $\omega_R$ and $\omega_f$ listed in Appendix B.

\par In FIG.(5), assuming the plot obtained for $\delta_R \approx1.5mm$ with $R_f \to 0$, 
we include the predictions for $\delta^c_R$ obtained from Eq.(\ref{eqs}). The points $({\bf\color{red}\blacktriangle},{\bf\color{red}\vartriangle},{\bf\color{red}\blacktriangledown})$ in this figure represent the determination of $\delta^c_R$ for the spherical floater with a diameter of $D=7mm$. The red dotted curve was obtained by setting $\delta^c_R = 0.26mm$.

\par   The $\delta^c_R$ values obtained are in good agreement with those observed for spherical floaters, as depicted in Figs. 4a-4c. However, it is crucial to consider that dynamic trapping introduces the effect of dynamic drag. This drag can lead to a deviation between the predictions of Eq.(\ref{eqs}) (derived under ideal conditions) and the experimentally observed $\delta^c_R$   values. A direct comparison reveals a deviation of approximately $\sim O(20\%)$, which is directly attributable to drag effects. Notably, this magnitude of deviation is consistent with that obtained in previous experiments carried out by the authors\cite{ref3}.

\par  Finally, in Tables (B5-B6) present results for cases where $R_f > R_{f}(min)$ . In these case, the $U_f(z_f)$ curves  which describe the trapping, are characterized by  $\delta^c_R < dR_f /5$, corresponding to the region between the blue and red dash-dotted curves shown in FIG.(4),  
for a given rotor configuration. The observed increase in $R_f$  for example,
compared to the data in Table B2 for distances of the same order  arises from differences in the rotor configuration, specifically the lateral displacement employed.

\par In the upper left corner of FIGs.(4a)-(4b), it can be observed that the rotor frequency required for trapping the same floater for the cases where $\delta_R \approx 1mm$ and $\delta_R \approx 1.5mm$,  are $\omega_R =12,330\, rpm$ and $\omega_R =8,946\, rpm$,respectively. 
As discussed previously, in this type of device,  angular momentum is transferred from the rotor to the floater.  For example, in the case where  $\delta_R \approx 1mm$ Table B6 shows that
the respective increase in $R_f$  results from the greater transfer of angular momentum from the rotor.

\par For the results presented in Table B5, where the lateral displacement is given by $\delta_R \approx 1. 75$mm, a comparison can be made with the data from Table B1. Since the energy available for trapping the floats is proportional to $\delta_R$, for this rotor configuration we have more energy available, so the trapping for $r_f  \approx (9.6-9.7) $mm occurs  at a lower rotation speed $\omega_R \approx^{^{\!\!\!\!\!B5}} O(11,100)$ rpm when compared to the trapping observed for $\delta_R \approx 1.5 $mm, where $\omega_R \approx^{^{\!\!\!\!\!B1}} O(12,500)$ rpm; the  indices $(B5,B1)$ refer to the respective tables.

\par Therefore, from the discussion raised in the above paragraphs, it is possible to verify that the increase in lateral displacement $\delta_R$, when we compare different configurations $\delta^A_R >  \delta^B_R$, is in principle, responsible for the magnetic trapping of the float observed at rotor speeds of the order of $O(10,000)rpm$ for the cases reported in FIGs.(4b),(4c),(4e) and (4f). In addition, by  $\omega^{^{\!B5}}_R < \omega^{^{\!B1}}_R $ discussed previously. Thus, an important question arises "Is there an upper limit for $\delta_R = \delta^{max}_{R}$,
beyond which the effect of magnetic trapping ceases to occur?".

\par Having established the dependence of $\omega_R$ , necessary for trapping on the choice attributed to $\delta_R$, we tested several variations specifically $\delta_R = 0.7h$ and $\delta_R = h/2$ maintaining a constant magnet geometry (a cylinder with $h=d=5mm$). We observed that when $\delta_R$ exceeded $h/2$, the trapping effect ceased to occur.

\par The performed set of measurements suggest the existence of an upper limit on $\delta_R$, which apparently corresponds to $\delta^{max}_R \approx 2.5mm$ for the geometry of the magnet coupled to the rotor considered in this case. In the following subsection we will discuss the theoretical possibility of an upper bound on $\delta_R$.

\subsection{The condition on the maximum  allowed  value of lateral displacement  }

\begin{figure}[t]
    \centering
\includegraphics[scale=0.6]{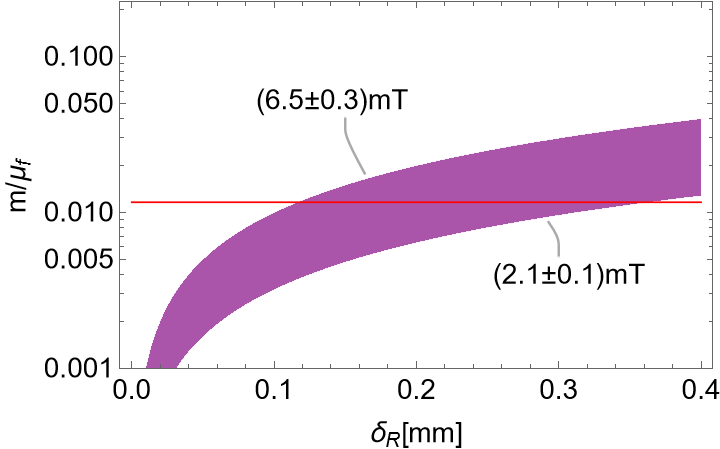}
    \caption{ The contextualization of the curves behavior are described in the text.}
    \label{fig2}
\end{figure}

\par   In our experiments, the magnet coupled to the rotor consisted of a cylindrical magnet with height and diameter $d=h=5mm$, assuming that the magnetization of the  cylinder is uniform,  the magnetic field at the center of the magnet can be expressed as:
\begin{equation}
  B_0 = \frac{\mu_0 M_R}{2},  
  \label{eqb0cil}
\end{equation}
\noindent where $M_R$ corresponds to the magnetization of the magnet. The magnetic moment $\mu_R$ can be written as $\mu_R=M_RV$, where  $V$ is the volume of the magnet considered. Consequently, 
 the magnetic field at the center of this cylinder can be estimated from the equation:
\begin{equation}
  B_0 = \frac{\mu_0 \mu_R}{2h^3}. 
  \label{eqB0}
\end{equation}
\par In Ref.\cite{ref8},  the behavior of the magnetic field $B(z)$ of a neodymium magnet with cylindrical geometry along its central axis, is described as
\begin{equation}
 B(z) = \frac{B_0}{2}\left(\frac{\frac{h}{2} - z}{\sqrt{\frac{h^2}{4} + (\frac{h}{2} - z)^2 } } +\frac{\frac{h}{2} + z}{\sqrt{\frac{h^2}{4} + (\frac{h}{2} + z)^2 } } \right),
    \label{eqBcil}    
\end{equation}
\noindent therefore, considering Eqs.(\ref{eqB0}) and (\ref{eqBcil}), the field along the central axis of the magnet coupled to the rotor can be characterized by $B(z) = B_0f(z)$. 

\par To investigate the existence of a possible upper bound for $\delta_R$, we revisit (\ref{eq7})  and reformulate it as

\begin{equation}
 \frac{F_{dip}r_f^3}{\mu_0\mu_f\mu_R} = \frac{3}{4\pi r_f}\left(\frac{4\delta_R}{r_f} -  R_f\right).
 \label{eqR0}
\end{equation}

\par  Considering a series of measurements conducted under the condition  $R_f = 0$ (see  FIG.(4)), we identified that the critical value for the lateral displacement $\delta^c_R(0)$ occurs within the region in which $z_f \approx 4mm$ or $r_f=r_{f(max)}\approx O(10)mm \approx 2h$. Then,  the magnetic trapping in this region can be characterized by

\begin{equation}
 \frac{2hmg}{\mu_fB_0} \approx \frac{3}{16\pi}\left(\frac{2\delta_R}{h}\right),
 \label{eqRat}
\end{equation}

\noindent where in obtaining the above equation, in addition to considering Eqs.(\ref{eqB0}) and (\ref{eqR0}) we also take into account that under the trapping condition the system is in dynamical equilibrium, allowing us to identify $\delta_R \approx \delta^c_R(0)$.

\par In Eq.(\ref{eqRat}),  the radius $m/\mu_f$ can be determined knowing the intensity of the rotor's magnetic field at the trapping point. Incorporating this into Equation Eq.(\ref{eqRat}) , alongside Equation (\ref{eqBcil}), and assuming the system is in equilibrium (utilizing Newton’s
third law), we can write Equation (\ref{eqRat}) as

\begin{figure}[t]
    \centering
\includegraphics[scale=0.35]{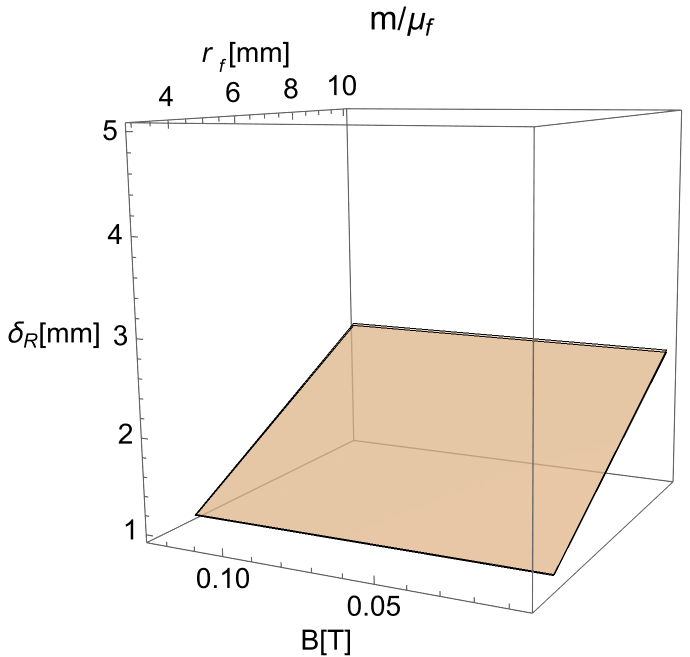}
    \caption{ 3D contour plot obtained from Eq.(\ref{eqRat234}).}
    \label{figxx}
\end{figure}

\begin{equation}
 \frac{2hmg}{\mu_fB(r_f)} \approx \frac{3}{16\pi f(r_f)}\left(\frac{2\delta_R}{h}\right). 
 \label{eqRat2}
\end{equation}
\par Now considering the direct measurement $B(r_f)$ as input, we can  infer from Eq.(\ref{eqRat2}) the behavior of the ratio $m/\mu_f$ at the point trapping is observed. 

\par In FIG.(6) we show how the radius $m/\mu_f$ varies as a function of $\delta_R$.
The region shaded in purple delimits the results obtained from  direct measurements of the magnetic field at a distance $r_f=10mm$, with the minimum and maximum values  $B_{min} =(2.1 \pm 0.1)mT $ and $B_{max} =(6.5 \pm 0.3)mT$, respectively. In this figure, the solid red line corresponds to the direct determination of $m/\mu_f = 0.01158 Kg/A.m^2$. From this plot,  we find $0.35mm \gtrsim  \delta^c_R >  0.1mm$ which agrees with the results found in subsection 5.1.

\par Eq.(\ref{eqRat2}) was derived considering the particular case in which $R_f = 0$, which defines the minimum energy configuration $E_f(min) = U(0)_{grav}$ for the observation of the magnetic trapping of spherical, cylindrical and cubic floaters depicted  in FIG.(4). 
When we observe the magnetic trapping in which $R_f > 0$ as illustrated in FIG.(5), it should be noted  that $E_f > E_f(min)$ , and $m/\mu_f$ is described by
\begin{equation}
 \frac{2r_fmg}{\mu_fB(r_f)} \approx \frac{3}{\pi f(r_f)}a(r_f)\left(b(r_f)\delta_R - R_f\right), 
 \label{eqRat234}
\end{equation}
\noindent where $a(r_f)=h^3/r^3_f$  and $b(r_f)= 4/r_f$. Measurements of $B(r_f)$ and $r_f$ allow the identification of $m/\mu_f(\delta_R)$, for a given specific rotor configuration defined by $\delta_R$. Assuming limiting case where $R_f=R_{f(max)} \to 1$ , by extrapolating from the 3D contour plot provided at FIG.(7) we can then  identify  for  $B_{min}(10mm) =(2.1 \pm 0.1)mT$ and   $B_{max}(3mm) =(0.125\pm 0.006)T$, where $ 3mm \lesssim  r_f \lesssim 10mm$ , that  
\begin{equation}
 \delta^{max}_R \lesssim 2.525 mm. 
 \label{eqRat3}
\end{equation}
\par From the upper bound for $\delta_R$ found above, we can finally write the prediction for the behavior of $m/\mu_f$ as given by Eq.(\ref{eqRat2}), which corresponds to the condition where the floater presents the minimum energy required for trapping,  in the form
\begin{equation}
 \frac{2hmg}{\mu_fB(r_f)} \approx \frac{15}{4}\left(\frac{\delta^c_R}{\delta^{max}_R}\right). 
 \label{eqRat22}
\end{equation}

\section{Conclusions}

\par   In Refs.\cite{ref3}\cite{ref4}, the authors established the fundamental physical principles underlying the magnetic levitation observed in conventional Levitron systems. 
These works demonstrate that stable levitation necessitates a precise tuning of the ratio $m_/\mu_f$, which ensures the system achieves a potential minimum at the trapping point.

\par  Building upon this established framework and considering the results from subsections V-1 to V-5, this paper extends these principles to a "dynamical Levitron" \cite{ref2}\cite{ref1}. The main difference between these two systems, as previously discussed, is that in the dynamical Levitron, the ratio $m_/\mu_f$ becomes dependent on the lateral displacement $\delta_R$.   Consequently,  the adjustment required for magnetic trapping is not governed by the rotation speed of the floater, but rather by the appropriate choice of $\delta_R$. This distinct characteristic enables trapping configurations where the floater can remain at rest, as shown in the videos available at\cite{videos1}.
  
\par By considering an effective model, we found that the observed levitation effect can be explained within the context of the discussion presented in\cite{ref3}\cite{ref4}, where for the potential energy to be a minimum at the trapping point,  we verify  the existence of a condition under the $\delta_R $ parameter.  As a consequence, as shown throughout sections IV and V, 
the occurrence of the magnetic trapping effect of the floater is primarily due to the condition that the lateral displacement $\delta_R > 0$, being observed the trapping when  $\delta_R  \gtrsim  \delta^c_R$. 

\par In Ref.\cite{ref1} the authors noted that the trapping effect is only observed,  if $\delta_R > 1mm$, where the results reported from simulations were performed with $\delta_R = 1.2mm$. As we discussed at the end of subsection B,  assuming  a naive dimensional analysis, we estimate that for  the experimental apparatus employed in \cite{ref1} that $\delta^c_R  \approx  O(1)mm$  which agrees with their observation. In this same paper, the authors report the observation of four different types of vibrational modes of the floater magnet, as a function of the frequency of the rotor magnet. 

\par In some situations involving magnetic trapping we also identified distinct vibrational modes. However, after noticing that the rotor and floater speeds became strongly correlated at the moment of floater trapping, we verified that instabilities and vibrations arising from the rotor ended up being transmitted at the moment of coupling to the floater, which were then removed with the appropriate mechanical isolation of the motor used in our experimental setup. 

\par Magnetic levitation has applications in several areas, such as Maglev trains, which use magnetic levitation to increase speed and reduce friction. However,  development costs for this type of train are quite high, making large-scale commercial implementation of this technology challenging. We believe that the type of magnetic levitation, resulting from the description provided in this work, has the potential to contribute to  the development of lower cost technologies. 
\vspace*{0.5cm}
\par {\bf Acknowledgments}
\par We would like to thank A. A. Natale for reading the
manuscript and for useful suggestions.
\vspace*{0.5cm}
\par {\bf  Data availability statement}
\par All data that support the findings of this study are included within the article (and any
 supplementary material in \cite{videos1}). It is important to highlight that \cite{videos1} refers to a repository containing all the video recordings of the experiments conducted in this study. 

\bibliographystyle{iopart-num}
\bibliography{references}

\providecommand{\newblock}{}
\begin{thebibliography}{1}
\expandafter\ifx\csname url\endcsname\relax
  \def\url#1{{\tt #1}}\fi
\expandafter\ifx\csname urlprefix\endcsname\relax\def\urlprefix{URL }\fi
\providecommand{\eprint}[2][]{\url{#2}}

\bibitem{ref3}
Simon M~D, Heflinger L~O and Ridgway S~L 1997 {\em Am. J. Phys\/} {\bf 65} 286--292

\bibitem{ref4}
Berry M~V 1996 {\em Proc. R. Soc.London, Ser. A\/} {\bf 452} 1207--1220

\bibitem{ref6}
Magnetics A 2025 {Neodymium Iron Boron Magnet Catalog} 210607 Arnold Magnetics https://www.arnoldmagnetics.com/wp-content/uploads/2019/06/Arnold-Neo-Catalog.pdf

\bibitem{ref2}
Ucar H 2021 {\em Symmetry\/} {\bf 13} 442

\bibitem{ref7}
Gans R~F, Jones T~B and Washizu M 1998 {\em Journal of Physics D: Applied Physics\/} {\bf 31} 671

\bibitem{ref1}
Hermansen J~M and et al 2023 {\em Phys.\ Rev. Apllied\/} {\bf 20} 044036--1

\bibitem{videos1}
Doff A and Szmoski R~M 2025 Videos and supplementary materials are available from the data repository at {http://paginapessoal.utfpr.edu.br/agomes} [A link to the corresponding YouTube channel containing all videos is also provided on this page.]

\bibitem{ref9}
Hermansen J~M, Durhuus F~L and Bjørk R 2025 {\em Applied Physics Letters\/} {\bf 127} 012402

\bibitem{ref8}
Doff A and Szmoski R~M 2023 {\em Rev. Bras. Ens. Fis.\/} {\bf 45} e20230181

\end{thebibliography}

\newpage
\appendix

   \section{Experimental Setup}

\begin{figure}[t]
    \centering
\hspace*{-0.5cm}\includegraphics[scale=0.5]{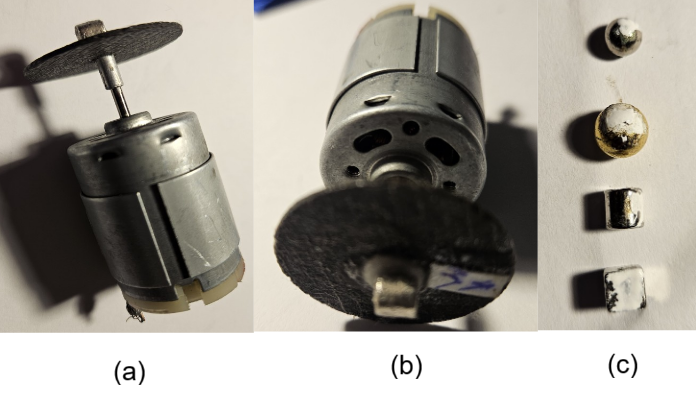}
 \vspace*{-0.1cm}
    \caption{The experimental setup: Motor used in the construction of the rotor and floats used. }
    \label{fig2}
\end{figure}

\par We have performed an experimental setup as follows.  As depicted in Figs.(A1-a),(A1-b) , a  Nd-Fe-B cylindrical magnet with diameter of $ d=5mm$ and height $h=5mm$, whose nominal remanence is $(1.17–1.22)T $,  was firmly attached to the motor shaft using super bonder glue. 
\par Coaxially with this magnet, a disk of radius $r=15mm$ was fixed to enable real-time measurement of the motor's rotational speed by a digital tachometer's laser sensor, as shown in
 Fig.(A2-a), that is used to determine the motor rotational speed in real time.

\par The motor identified in Fig.(A1), consists of a micro motor DC  whose operating specifications correspond to $V_{max}=24V$, $i_{max} = 1.21A$ and $\omega_{max} = 21,600rpm$. The floaters employed in the experiments, documented in the videos available in \cite{videos1} and illustrated in Fig. (A1-c)\footnote{Note that in the preparation of this photo, the magnets were placed under a small metal plate inserted below a sheet of white paper.}, were marked to facilitate the
 determination of $\omega_f$ using the motion-tracking software Tracker. 

\par To control the motor's speed, a stabilized power source, depicted in Fig. A2-b, was employed. This source allows for precise adjustment of voltage (V) and current (I), thereby controlling its rotational speed (($\omega$). As detailed in the tables provided in Appendix B, the motor rotational speeds required for the magnetic trapping of the float typically range from $\omega_f (10,400 - 13,700)rpm$. The intrinsic alternating magnetic field of the motor at a typical distance of the floater magnet is assumed to be negligible. 
 
\begin{figure}[b]
    \centering
 \vspace*{-0.6cm}
\hspace*{-0.5cm}\includegraphics[scale=0.4]{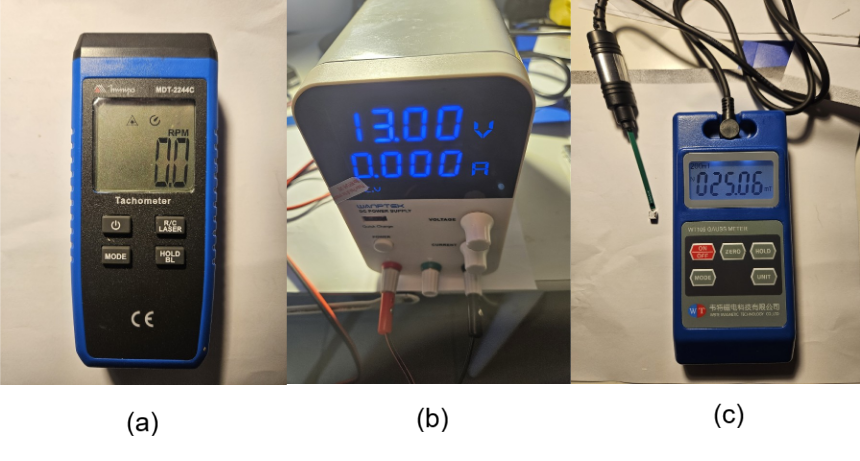}
 \vspace*{-0.1cm}
    \caption{The experimental setup: Devices used in the characterization of measurements. }
    \label{fig2}
\end{figure}
\newpage 

\begin{figure}[t]
    \centering
\hspace*{-0.7cm}\includegraphics[scale=0.6]{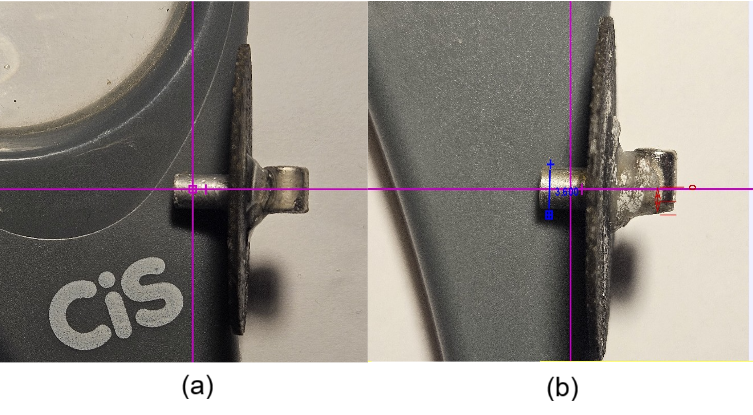}
 \vspace*{-0.1cm}
    \caption{The experimental setup: Lateral displacements $\delta_R = 1mm$ (a) and   $\delta_R = 1.75mm$ (b). }
    \label{fig2}
\end{figure}

\par  The  distinct configurations for lateral displacement $\delta_R$ that we consider in this work were incorporated in different arrangements similar to those shown in Fig.(A1). For illustration purposes, in Fig.(A3) we present the configuration in which $\delta_R=1mm$ , Fig.(A3-a) and $\delta_R=1.75mm$ in Fig.(A3-b).

\par  Figures A4-a and A4-b illustrate the methodology employed to obtain the tabulated results in Appendix B. During experiments, the dynamical behavior of the
floater magnet was recorded and these video recordings were subsequently post-processed using the motion-tracking software Tracker, in order to determine the position and orientation of the floater as a function of time.

\begin{figure}[b]
    \centering
\hspace*{-0.7cm}\includegraphics[scale=0.55]{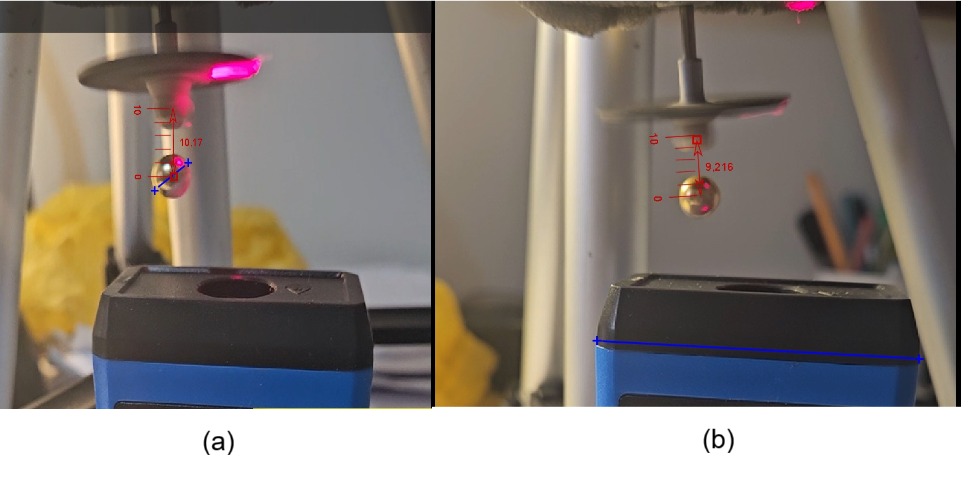}
 \vspace*{-0.1cm}
    \caption{The experimental setup:  High-resolution camera footage at 920 fps of the magnet levitating. }
    \label{fig2}
\end{figure}

\par  As we mentioned, the determination of the rotor(R) rotation speed, $\omega_R$, is obtained from the direct reading provided by the digital tachometer. The specific instrument used for these measurements is a Minipa Mdt-2244c digital photo tachometer. This device features a rotational speed range from  $\omega_{min} = 10rpm$ to $\omega_{max} = 99,999 rpm$,  and  the minimum and maximum distance range between the sensor and the rotating surface corresponds to
$(50-200)mm$. The accuracy in determining the rotational speed, for a reading falling within the range $(50-200)mm$ , corresponds to $\Delta \omega = \pm 0.04\%$. 
\newpage

\par  In turn, the floater's rotational speed, $\omega_f$, was determined through the analysis of high-resolution slow-motion videos using the Tracker software. These recordings were captured at 960 frames per second (fps) with a Samsung Galaxy S23 mobile phone.  The error in Tracker analysis varies significantly by application, but common sources of error include video quality (resolution, blur), calibration accuracy, object visibility, and the precision of manual or automatic point marking. 

\par There is no single "error" value for Tracker; instead, it depends on specific experimental conditions. Manually marking the center of an object in each frame in videos is a source of uncertainty, that can estimate as  $O(1)\%$,  and  can be assumed in the  our estimates  for $\omega_f$. 

\par  Finally, the digital Gaussmeter WT10A represented in Fig.(A2-c), was used to characterize the magnetic field measurements reported in section V-5, the magnetic field measurement range for this device corresponds to $(0-2000)mT$, presenting an accuracy of $\Delta B = \pm 5\%$. Considering the resolution of this device, the magnetic field measurements reported in subsection V-5 correspond to $B_{min} = (2.1 \pm 0.1)mT $ and $B_{max} = (6.5 \pm 0.3)mT$.

\newpage

 \section{Tables}

\begin{table}[htbp]
\caption{Experimental measurements obtained for $\omega_R(rpm)$ and $\omega_f(rpm)$ and $r_f(mm)$ for a cylindrical float with diameter and height of $5mm$ and $\delta_R = 1,5mm$. The measurements obtained for $\omega_{f}$ are the result of using the Tracker on the film obtained with a high-resolution camera (960 fps).The $\omega_R$ measurement was performed considering direct reading using a laser tachometer.}

\centering
\begin{tabular}{cccccc}
\textrm{Measure}&
\textrm{$\omega_R(rpm)$}&
\textrm{$\omega_f(rpm)$}&
\textrm{$R_f(rad)$}&
\textrm{$r_f(mm)$}&
\textrm{$\bar{R}_f(rad)$}\\
 \hline 
1 & 12,740 $\pm$ 5.1 & 1,538.5 $\pm$ 15.4 & 0.120 $\pm$ 0.0012 &  &     \\
2 & 12,725 $\pm$ 5.1 & 1,538.5 $\pm$ 15.4 & 0.120 $\pm$ 0.0012 &  &     \\ 
3 & 12,735 $\pm$ 5.1 & 1,450.2 $\pm$ 14.5 & 0.114 $\pm$ 0.0011 &  9.40 $\pm$ 0.094  &  0.12 $\pm$ 0.0012\\ 
4 & 12,734 $\pm$ 5.1 & 1,531.1 $\pm$ 15.3 & 0.120 $\pm$ 0.0012 &   &     \\ 
5 & 12,750 $\pm$ 5.1 & 1,576.3 $\pm$ 15.8 & 0.124 $\pm$ 0.0012 &  &     \\ 
6 & 12,760 $\pm$ 5.1 & 1,504.7 $\pm$ 15.0 & 0.117 $\pm$ 0.0012 &  &    \\
\hline
1 & 12,576 $\pm$ 5 & 1,395.3 $\pm$ 13.9& 0.111 $\pm$ 0.0011 &  &     \\
2 & 12,512 $\pm$ 5 & 1,371.4 $\pm$ 13.7& 0.104 $\pm$ 0.0010 &  &     \\ 
3 & 12,500 $\pm$ 5 & 1,474.6 $\pm$ 14.7& 0.118 $\pm$ 0.0012 &  9.60 $\pm$ 0.096  &  0.115 $\pm$ 0.0011\\ 
4 & 12,512 $\pm$ 5 & 1,500.0 $\pm$ 15.0& 0.120 $\pm$ 0.0012 &   &     \\ 
5 & 12,545 $\pm$ 5 & 1,461.2 $\pm$ 14.6& 0.116 $\pm$ 0.0012 &  &     \\ 
6 & 12,536 $\pm$ 5 & 1,504.0 $\pm$ 15.0& 0.120 $\pm$ 0.0012 &  &      \\
\hline 
1 & 12,225 $\pm$ 4.9 & 1,344.5 $\pm$ 13.4 & 0.110 $\pm$ 0.0011 &  &     \\
2 & 12,245 $\pm$ 4.9 & 1,389.3 $\pm$ 13.9 & 0.113 $\pm$ 0.0011 &  &     \\ 
3 & 12,254 $\pm$ 4.9 & 1,333.3 $\pm$ 13.3 & 0.109 $\pm$ 0.0010 &  10.08 $\pm$ 0.1  &  0.11 $\pm$ 0.0011\\ 
4 & 12,258 $\pm$ 4.9 & 1,344.5 $\pm$ 13.4 & 0.110 $\pm$ 0.0011 &   &     \\ 
5 & 12,266 $\pm$ 4.9 & 1,350.2 $\pm$ 13.5 & 0.110 $\pm$ 0.0011 &  &     \\ 
6 & 12,254 $\pm$ 4.9 & 1,359.8 $\pm$ 13.6 & 0.110 $\pm$ 0.0011&  &    \\
\hline 
\end{tabular} 
    
\end{table}

\begin{table}[htbp]
\caption{Experimental measurements obtained for $\omega_R(rpm)$ and $\omega_f(rpm)$ and $r_f(mm)$ for a spherical float with diameter of $7mm$ and $\delta_R = 1,5mm$. The measurements obtained for $\omega_{f}$ are the result of using the Tracker on the film obtained with a high-resolution camera (960 fps).The $\omega_R$ measurement was performed considering direct reading using a laser tachometer.}

\centering
\begin{tabular}{cccccc}
\textrm{Measure}&
\textrm{$\omega_R(rpm)$}&
\textrm{$\omega_f(rpm)$}&
\textrm{$R_f(rad)$}&
\textrm{$r_f(mm)$}&
\textrm{$\bar{R}_f(rad)$}\\
\hline 
1 & 10,843 $\pm$ 4.3 & 1,502.3 $\pm$ 15.0 & 0.138 $\pm$ 0.0014&  &     \\
2 & 10,855 $\pm$ 4.3 & 1,490.7 $\pm$ 14.9 & 0.137 $\pm$ 0.0014&  &     \\ 
3 & 10,837 $\pm$ 4.3 & 1,490.7 $\pm$ 14.9 & 0.137 $\pm$ 0.0014&  9.52$\pm$ 0.095&  0.134$\pm$ 0.0013\\ 
4 & 10,835 $\pm$ 4.3 & 1,413.8 $\pm$ 14.1 & 0.130 $\pm$ 0.0013&   &     \\ 
5 & 10,816 $\pm$ 4.3 & 1,521.4 $\pm$ 15.2 & 0.140 $\pm$ 0.0014&  &     \\ 
6 & 10,815 $\pm$ 4.3 & 1,322.0 $\pm$ 13.2 & 0.122 $\pm$ 0.0012&  &    \\
\hline
1 & 10,645 $\pm$ 4.2 & 1,346.4 $\pm$ 13.4 & 0.126 $\pm$ 0.0013&  &     \\
2 & 10,630 $\pm$ 4.2 & 1,350.2 $\pm$ 13.5 & 0.127 $\pm$ 0.0013&  &     \\ 
3 & 10,616 $\pm$ 4.2 & 1,320.5 $\pm$ 13.2 & 0.124 $\pm$ 0.0012&  9.74$\pm$ 0.097 &  0.125$\pm$ 0.0012\\ 
4 & 10,620 $\pm$ 4.2 & 1,342.6 $\pm$ 13.4 & 0.126 $\pm$ 0.0013&   &     \\ 
5 & 10,585 $\pm$ 4.2 & 1,315.0 $\pm$ 13.1 & 0.124 $\pm$ 0.0012&  &     \\ 
6 & 10,613 $\pm$ 4.2 & 1,316.9 $\pm$ 13.2 & 0.125 $\pm$ 0.0012&  &      \\
\hline 
1 & 10,460 $\pm$ 4.2 & 1,286.9 $\pm$ 12.9 & 0.123 $\pm$ 0.0012&  &     \\
2 & 10,430 $\pm$ 4.2 & 1,269.0 $\pm$ 12.7 & 0.122 $\pm$ 0.0012&  &     \\ 
3 & 10,425 $\pm$ 4.2 & 1,250.6 $\pm$ 12.5 & 0.120 $\pm$ 0.0012&  9.94$\pm$ 0.1 &  0.12$\pm$ 0.0012\\ 
4 & 10,415 $\pm$ 4.2 & 1,232.7 $\pm$ 12.3 & 0.118 $\pm$ 0.0012&   &     \\ 
5 & 10,426 $\pm$ 4.2 & 1,269.4 $\pm$ 12.7 & 0.122 $\pm$ 0.0012&  &     \\ 
6 & 10,422 $\pm$ 4.2 & 1,222.9 $\pm$ 12.23 & 0.117 $\pm$ 0.0012&  &      \\
\hline 
\end{tabular}
\end{table}

\begin{table}[t]
\caption{Experimental measurements obtained for $\omega_R(rpm)$ and $\omega_f(rpm)$ and $r_f(mm)$ for a spherical float with diameter of $4mm$ and $\delta_R = 1,5mm$. The measurements obtained for $\omega_{f}$ are the result of using the Tracker on the film obtained with a high-resolution camera (960 fps).The $\omega_R$ measurement was performed considering direct reading using a laser tachometer.}

\centering
\begin{tabular}{cccccc}
\textrm{Measure}&
\textrm{$\omega_R(rpm)$}&
\textrm{$\omega_f(rpm)$}&
\textrm{$R_f(rad)$}&
\textrm{$r_f(mm)$}&
\textrm{$\bar{R}_f(rad)$}\\
\hline 
1 & 10,714 $\pm$ 4.3& 1,167.9  $\pm$ 11.2& 0.109 $\pm$ 0.0011&  &     \\
2 & 10,724 $\pm$ 4.3& 1,112.4  $\pm$ 11.1& 0.104 $\pm$ 0.0010&  &     \\ 
3 & 10,704 $\pm$ 4.3& 1,151.1  $\pm$ 11.5& 0.107 $\pm$ 0.0011&  11.26$\pm$ 0.11 &  0.11$\pm$ 0.0011\\ 
4 & 10,708 $\pm$ 4.3& 1,188.1  $\pm$ 11.9& 0.110 $\pm$ 0.0011&   &     \\ 
5 & 10,648 $\pm$ 4.3& 1,125.4  $\pm$ 11.2& 0.106 $\pm$ 0.0011&  &     \\ 
6 & 10,708 $\pm$ 4.3& 1,155.2  $\pm$ 11.5& 0.108 $\pm$ 0.0011&  &    \\
\hline
1 & 11,019 $\pm$ 4.4& 1,452.3 $\pm$ 14.5& 0.132  $\pm$ 0.0013&  &     \\
2 & 11,050 $\pm$ 4.4& 1,488.4 $\pm$ 14.9& 0.134  $\pm$ 0.0013&  &     \\ 
3 & 11,045 $\pm$ 4.4& 1,465.6 $\pm$ 14.6& 0.132  $\pm$ 0.0013&  10.22 $\pm$ 0.1 &  0.132 $\pm$ 0.0013\\ 
4 & 11,024 $\pm$ 4.4& 1,447.9 $\pm$ 14.5& 0.131  $\pm$ 0.0013&   &     \\ 
5 & 11,036 $\pm$ 4.4& 1,450.2 $\pm$ 14.5& 0.131  $\pm$ 0.0013&  &     \\ 
6 & 11,035 $\pm$ 4.4& 1,461.2 $\pm$ 14.6& 0.132  $\pm$ 0.0013&  &      \\
\hline 
1 & 11,424 $\pm$ 4.6& 1,561.0 $\pm$ 15.6& 0.137  $\pm$ 0.0014&  &     \\
2 & 11,429 $\pm$ 4.6& 1,502.3 $\pm$ 15.0& 0.131  $\pm$ 0.0013&  &     \\ 
3 & 11,402 $\pm$ 4.6& 1,521.4 $\pm$ 15.2& 0.133  $\pm$ 0.0013&  9.76 $\pm$ 0.097 &  0.134 $\pm$ 0.0013\\ 
4 & 11,400 $\pm$ 4.6& 1,521.4 $\pm$ 15.2& 0.133  $\pm$ 0.0013&   &     \\ 
5 & 11,386 $\pm$ 4.6& 1,543.4 $\pm$ 15.4& 0.135  $\pm$ 0.0013&  &     \\ 
6 & 11,420 $\pm$ 4.6& 1,553.4 $\pm$ 15.5& 0.136  $\pm$ 0.0014&  &     \\
\hline 
\end{tabular}
\end{table}
\begin{table}[b]
\caption{Experimental measurements obtained for $\omega_R(rpm)$ and $\omega_f(rpm)$ and $r_f(mm)$ for a spherical float with diameter of $7mm$ and $\delta_R = 1,75 mm$. The measurements obtained for $\omega_{f}$ are the result of using the Tracker on the film obtained with a high-resolution camera (960 fps).The $\omega_R$ measurement was performed considering direct reading using a laser tachometer.}

\centering
\begin{tabular}{cccccc}
\textrm{Measure}&
\textrm{$\omega_R(rpm)$}&
\textrm{$\omega_f(rpm)$}&
\textrm{$R_f(rad)$}&
\textrm{$r_f(mm)$}&
\textrm{$\bar{R}_f(rad)$}\\
\hline 
1 & 11,943  $\pm$ 4.8& 1,801.12 $\pm$ 18.0& 0.150  $\pm$ 0.0015&  &     \\
2 & 11,950  $\pm$ 4.8& 1,777.80 $\pm$ 17.8& 0.149  $\pm$ 0.0015&  &     \\ 
3 & 11,944  $\pm$ 4.8& 1,708.20 $\pm$ 17.1& 0.143  $\pm$ 0.0014&  9.22 $\pm$ 0.092 &  0.145 $\pm$ 0.0014\\ 
4 & 11,964  $\pm$ 4.8& 1,693.12 $\pm$ 16.9& 0.141  $\pm$ 0.0014&   &     \\ 
5 & 11,964  $\pm$ 4.8& 1,684.20 $\pm$ 16.8& 0.140  $\pm$ 0.0014&  &     \\ 
6 & 11,958  $\pm$ 4.8& 1,751.80 $\pm$ 17.5& 0.146  $\pm$ 0.0015&  &    \\
\hline
1 & 11,505 $\pm$ 4.6& 1,497.66 $\pm$ 15.0& 0.130  $\pm$ 0.0013&  &     \\
2 & 11,509 $\pm$ 4.6& 1,509.40 $\pm$ 15.1& 0.131  $\pm$ 0.0013&  &     \\ 
3 & 11,504 $\pm$ 4.6& 1,487.48 $\pm$ 14.9& 0.129  $\pm$ 0.0013&  9.60 $\pm$ 0.096 &  0.129 $\pm$ 0.0013\\ 
4 & 11,511 $\pm$ 4.6& 1,472.40 $\pm$ 14.7& 0.128  $\pm$ 0.0013&   &     \\ 
5 & 11,516 $\pm$ 4.6& 1,486.10 $\pm$ 14.9& 0.129  $\pm$ 0.0013&  &     \\ 
6 & 11,524 $\pm$ 4.6& 1,486.10 $\pm$ 14.9& 0.129  $\pm$ 0.0013&  &    \\
\hline 
1 & 11,025  $\pm$ 4.4& 1,241.70 $\pm$ 12.4& 0.113  $\pm$ 0.0011&  &     \\
2 & 11,040  $\pm$ 4.4& 1,200.00 $\pm$ 12.0& 0.109  $\pm$ 0.0011&  &     \\ 
3 & 11,044  $\pm$ 4.4& 1,279.20 $\pm$ 12.8& 0.116  $\pm$ 0.0012& 10.03 $\pm$ 0.1 &  0.113 $\pm$ 0.0011\\ 
4 & 11,054  $\pm$ 4.4& 1,273.20 $\pm$ 12.7& 0.115  $\pm$ 0.0011&   &     \\ 
5 & 11,045  $\pm$ 4.4& 1,218.30 $\pm$ 12.2& 0.110  $\pm$ 0.0011&  &     \\ 
6 & 11,062  $\pm$ 4.4& 1,297.30 $\pm$ 13.0& 0.117  $\pm$ 0.0012&  &   \\ 
\hline 
\end{tabular}
\end{table}

\begin{table}[t]
\caption{Experimental measurements obtained for $\omega_R(rpm)$ and $\omega_f(rpm)$ and $r_f(mm)$ for a cylindrical float with diameter and height of $5mm$ and $\delta_R = 1,75mm$. The measurements obtained for $\omega_{f}$ are the result of using the Tracker on the film obtained with a high-resolution camera (960 fps).The $\omega_R$ measurement was performed considering direct reading using a laser tachometer.}

\centering
\begin{tabular}{cccccc}
\textrm{Measure}&
\textrm{$\omega_R(rpm)$}&
\textrm{$\omega_f(rpm)$}&
\textrm{$R_f(rad)$}&
\textrm{$r_f(mm)$}&
\textrm{$\bar{R}_f(rad)$}\\
\hline 
1 & 11,575  $\pm$ 4.6& 2,519.70 $\pm$ 25.2& 0.217  $\pm$ 0.0021&  &     \\
2 & 11,559  $\pm$ 4.6& 2,594.60 $\pm$ 25.9& 0.224  $\pm$ 0.0022&  &     \\ 
3 & 11,574  $\pm$ 4.6& 2,537.73 $\pm$ 25.4& 0.220  $\pm$ 0.0022&  9.4 $\pm$ 0.094 &  0.224 $\pm$ 0.0022\\ 
4 & 11,565  $\pm$ 4.6& 2,630.10 $\pm$ 26.3& 0.223  $\pm$ 0.0022&   &     \\ 
5 & 11,294  $\pm$ 4.6& 2,580.64 $\pm$ 25.8& 0.230  $\pm$ 0.0023&  &     \\ 
6 & 11,577  $\pm$ 4.6& 2,666.70 $\pm$ 26.7& 0.230  $\pm$ 0.0023&  &    \\
\hline
1 & 11,194 $\pm$ 4.5& 1,726.60 $\pm$ 17.3& 0.154  $\pm$ 0.0015&  &     \\
2 & 11,175 $\pm$ 4.5& 1,748.63 $\pm$ 17.5& 0.156  $\pm$ 0.0016&  &     \\ 
3 & 11,163 $\pm$ 4.5& 1,804.50 $\pm$ 18.0& 0.160  $\pm$ 0.0016&  9.70 $\pm$ 0.097 &  0.162 $\pm$ 0.0016\\ 
4 & 11,133 $\pm$ 4.5& 1,867.70 $\pm$ 18.7& 0.168  $\pm$ 0.0017&   &     \\ 
5 & 11,142 $\pm$ 4.5& 1,867.70 $\pm$ 18.7& 0.168  $\pm$ 0.0017&  &     \\ 
6 & 11,138 $\pm$ 4.5& 1,867.70 $\pm$ 18.7& 0.168  $\pm$ 0.0017&  &   \\ 
\hline 
\end{tabular}
\end{table}

\begin{table}[h]
\caption{Experimental measurements obtained for $\omega_R(rpm)$ and $\omega_f(rpm)$ and $r_f(mm)$ for a spherical float with diameter of $7mm$ and $\delta_R = 1,00 mm$. The measurements obtained for $\omega_{f}$ are the result of using the Tracker on the film obtained with a high-resolution camera (960 fps).The $\omega_R$ measurement was performed considering direct reading using a laser tachometer.}

\centering
\begin{tabular}{cccccc}
\textrm{Measure}&
\textrm{$\omega_R(rpm)$}&
\textrm{$\omega_f(rpm)$}&
\textrm{$R_f(rad)$}&
\textrm{$r_f(mm)$}&
\textrm{$\bar{R}_f(rad)$}\\
\hline 
1 & 12,885  $\pm$ 5.1& 2,953.85  $\pm$ 29.5& 0.230  $\pm$ 0.0023&  &     \\
2 & 12,884  $\pm$ 5.1& 3,127.03  $\pm$ 31.3& 0.243  $\pm$ 0.0024&  &     \\ 
3 & 12,870  $\pm$ 5.1& 3,127.03  $\pm$ 31.3& 0.242  $\pm$ 0.0024&  9.70 $\pm$ 0.097 &  0.242 $\pm$ 0.0024\\ 
4 & 12,884  $\pm$ 5.1& 3,096.77  $\pm$ 31.0& 0.240  $\pm$ 0.0024&  &     \\ 
5 & 12,890  $\pm$ 5.1& 3,200.00  $\pm$ 32.0& 0.248  $\pm$ 0.0025&  &     \\ 
6 & 12,876  $\pm$ 5.1& 3,254.00  $\pm$ 32.5& 0.252  $\pm$ 0.0025&  &    \\
\hline
1 & 13,627  $\pm$ 5.4& 4,173.91 $\pm$ 41.7& 0.306  $\pm$ 0.0030&  &     \\
2 & 13,612  $\pm$ 5.4& 4,465.12 $\pm$ 44.6& 0.328  $\pm$ 0.0033&  &     \\ 
3 & 13,616  $\pm$ 5.4& 4,424.00 $\pm$ 44.2& 0.325  $\pm$ 0.0032&  9.22 $\pm$ 0.092 &  0.320 $\pm$ 0.0032\\ 
4 & 13,618  $\pm$ 5.4& 4,465.12 $\pm$ 44.6& 0.328  $\pm$ 0.0033&  &     \\ 
5 & 13,628  $\pm$ 5.4& 4,240.00 $\pm$ 42.4& 0.311  $\pm$ 0.0031&  &     \\ 
6 & 13,616  $\pm$ 5.4& 4,403.70 $\pm$ 44.0& 0.323  $\pm$ 0.0032&  &    \\
\hline 
\end{tabular}
\end{table}

\end{document}